\documentclass[usenatbib, useAMS]{mnras}

\usepackage[modulo]{lineno} % the modulo causes only one in 5 to be printed
\usepackage{xfrac} %  allows for nice fractions in text 
\usepackage[svgnames]{xcolor}  %
\usepackage{amsmath}  % provides "split" to split equation over two lines
\usepackage{lastpage}  % need a \label{lastpage} somewhere at end

\newcommand{\arcdeg}{\degr}  % because I'm used to \arcdeg
\newcommand{\uv}{\mbox{$u$-$v$}}

\newcommand{\kms}{\mbox{km s$^{-1}$}}

\newcommand{\Jb}{\mbox Jy~beam$^{-1}$}
\newcommand{\muJb}{\mbox{$\mu$Jy~beam$^{-1}$}}

\newcommand{\gtrsim}{\mbox{\raisebox{-0.3em}{$\stackrel{\textstyle >}{\sim}$}}}

\newcommand{\thout}{\mbox{$\theta_{\rm o}$}}
\newcommand{\thin}{\mbox{$\theta_{\rm i}$}}

\newcommand{\HII}{\mbox{H{\small\rmfamily{II}}}}
\newcommand{\roi}{\mbox{$\mathcal R_\textnormal{o/i}$}}  % textnormal renders
\newcommand{\thfl}{\mbox{$\theta_{\rm90\%\;flux}$}}

\newcommand{\tablenotemark}[1]{$^{\mathrm #1}$}
\newcommand{\tablenotetext}[2]{\noindent$^{\mathrm #1}$ #2\\}
\newcommand{\phn}{\phantom{1}}

\title[SN 1996cr VLBI]{The Bright Supernova 1996cr in the Circinus Galaxy Imaged with
  VLBI: Shell Structure with Complex Evolution}

\author[Bietenholz et al]{Michael F. Bietenholz$^{1,2}$,
Norbert Bartel$^1$,  
Franz E. Bauer$^{3,4,5}$, 
Vikram V. Dwarkadas$^{6}$, \and
Leon Mtshweni$^{7}$,
Carlos Orquera-Rojas$^{3,4}$,
Simon Ellingsen$^{8}$,
Shinji Horiuchi$^{9}$,\and
and Anastasios Tzioumis$^{10}$
\\ 
$^1$Department of Physics and Astronomy, York University, Toronto,
M3J~1P3, Ontario, Canada \\
$^2$SARAO/Hartebeesthoek Radio Astronomy Observatory, PO Box 443, Krugersdorp,
1740, South Africa \\
$^3$Facultad de Física, Instituto de Astrof\'isica and Centro de Astroingenier\'ia, Pontificia Universidad Cat\'olica de Chile, Casilla 306,\\ \phantom{'x'}Santiago 22, Chile\\
$^4$Millennium Institute of Astrophysics (MAS), Nuncio Monseñor S\'otero Sanz 100, Providencia, Santiago, Chile \\
$^5$Space Science Institute,~4750 Walnut Street, Suite 205, Boulder, CO 80301, USA \\
$^6$Department of Astronomy and Astrophysics, University of Chicago, 5640 S Ellis Avenue, Chicago, IL 60637, USA \\
$^7$Department of Physics, University of Pretoria, Hatfield, Pretoria, 0028, South Africa \\
$^8$School of Natural Sciences, University of Tasmania, Private Bag 37, Hobart, TAS 7001, Australia \\
$^9$CSIRO Space \& Astronomy, Canberra Deep Space Communications Complex, PO Box 1035, Tuggeranong, ACT 2901, Australia \\
$^{10}$Australia Telescope National Facility, CSIRO, PO Box 76, Epping, NSW 1710, Australia
}

\begin{document}
%\date{Version 4.2, \today}
\date{\em Accepted for publication in MNRAS}

\pagerange{\pageref
{firstpage}--\pageref{lastpage}} \pubyear{2022}
\maketitle
\label{firstpage}

\begin{abstract}
  % 5370 = MJD 55338 = 22-05-2010
  We present broadband radio flux-density measurements supernova (SN)
  1996cr, made with MeerKAT, ATCA and ALMA, and images made from very
  long baseline interferometry (VLBI) observations with the Australian
  Long Baseline Array.  The spectral energy distribution of SN~1996cr
  in 2020, at age, $t \sim$8700~d, is a power-law, with flux density,
  $S \propto \nu^{-0.588 \pm 0.011}$ between 1 and 34 GHz, but may
  steepen at $>35$~GHz.  The spectrum has flattened since $t = 5370$~d
  (2010).  Also since $t = 5370$~d, the flux density has declined
  rapidly, with $S_{\rm 9 \, GHz} \propto t^{-2.9}$\@.  The VLBI image
  at $t = 8859$~d shows an approximately circular structure, with a
  central minimum reminiscent of an optically-thin spherical shell of
  emission.  For a distance of 3.7~Mpc, the average outer radius of
  the radio emission at $t = 8859$~d was $(5.1 \pm 0.3) \times
  10^{17}$~cm, and SN~1996cr has been expanding with a velocity of
  $4650 \pm 1060$~\kms\ between $t=4307$ and 8859~d.  It must have
  undergone considerable deceleration before $t = 4307$~d.  Deviations
  from a circular shell structure in the image suggest a range of
  velocities up to $\sim$7000~\kms, and hint at the presence of a
  ring- or equatorial-belt-like structure rather than a complete
  spherical shell.
  % emacs says 223 words
\end{abstract}


\section{Introduction}
\label{sintro}
 
Supernova (SN)~1996cr, in the nearby Circinus Galaxy had the highest
radio flux density
ever observed for optically-identified radio supernova \citep[163 mJy
  at 8.3 GHz;][]{Bauer2007c, Bauer+2008}, as well as being one of only
a handful of supernovae which can still be observed in radio more than 20~yr
after the explosion.
It was seen in 2001 in X-rays
\citep{Sambruna+2001}, and subsequently identified as a possible SN
\citep{Bauer+2001}, but only firmly identified as a SN in 2008
\citep{Bauer+2008}.  It was not identified till well after the
explosion, so the explosion date, $t_0$, is not accurately known. We
take $t_0=$ 1995 Sep.\ 7 (MJD = 49968), which is the midpoint of the
range given in \citet{Bauer+2008}, and all our times, $t$, are with
respect to this $t_0$.
We take the Circinus Galaxy and SN~1996cr to be at a distance
$D \sim 3.7$~Mpc,\footnote{We average the distance derived from the
redshift using the NASA/IPAC Extragalactic Database (NED;
\url{https://ned.ipac.caltech.edu}), after correction for infall to
Virgo, the Great Attractor and Shapley supercluster and use $H_0 =
67.4$~\kms~Mpc$^{1}$ \citep{Planck+2018vi}, which distance is $3.16
\pm 0.22$~Mpc, and the redshift-independent value listed on NED of
4.2~Mpc to arrive at 3.7 Mpc.  However we note that there is some
disagreement on the distance to Circinus, see discussion in
\citet{Mondal+2021}, with values as high as 4.2 Mpc
\citep[e.g.][]{Freeman+1977} and as low as 2.6 Mpc
\citep{Rozanska+2018} being used in the literature.}
and we indicate the dependence of our derived quantities on
$D$, which is only known to about 25\% accuracy.

SN~1996cr was classified as a Type IIn supernova in 2007, around a
decade after the explosion (\citealt{Bauer+2008}; see also
\citealt{Ransome+2021}),
but there are no observations constraining its type at early
times.\footnote{Some SNe, e.g.\ SN~2014C
\citep{Milisavljevic+2015_SN2014C}, initially show normal Type II
spectra, and only after some time evolve to reveal IIn
characteristics.}  Type IIn supernovae are characterized by relatively
narrow H features superimposed on the broader emission lines, and are
associated with strong interaction with a dense circumstellar medium
\citep[CSM; e.g.][]{Fransson+2014, Smith2014} produced by winds and
outflows from the progenitors before the explosions. Type IIn SNe (for
example SN~1986J and SN~1988Z) tend to exhibit stronger and
longer-lasting radio emission than other Types of SNe
\citep{RadioLumFn}, as well as often having stronger X-ray emission at
late times \citep[e.g.][]{DwarkadasG2012}.

SN~1996cr has remained unusually bright in both radio and X-rays for
more than two decades, and is one of only a small number of long-lived
core-collapse SNe (CCSNe) whose evolution can be studied over more
than a decade.
It's close distance of 3.7~Mpc makes it easier to observe than most
other Type II SNe.  SN~1996cr showed a rapid rise in radio emission
around 1~yr after the explosion and then a plateau lasting for several
more years \citep{Bauer+2008, Meunier+2013}.  The X-rays showed a more
gradual increase lasting till $\sim$10~yr after the explosion
\citep{Quirola-Vasquez+2019}.

SN~1996cr's rise in radio and X-rays is attributed to the SN shock
interacting with a dense region of CSM \citep{Bauer+2008,
  DwarkadasDB2010}. SN~1996cr first exploded in a lower-density region,
but then $1 \sim 2$~yr after the explosion, the forward shock
interacted with a region of dense CSM,
%C  Leave this as is.  We don't know dense CSM *is* a shell, we're mostly
%C  talking as if it is *not* a shell but rather waist of peanut-shape
%C  It was Bauer & DwarkadasDB who described it as a shell.
which is characterized by \citet{Bauer+2008} and
\citet{DwarkadasDB2010} as a shell formed during an episode of
mass-loss from the progenitor.
The inner radius of the dense CSM has been taken
to be $\sim 10^{17}$ cm based largely on modelling of the emission at
various wavebands \citep{DwarkadasDB2010, DeweyBD2011,
  Quirola-Vasquez+2019} and constrained by the first
VLBI measurement \citep{Bauer+2008}.
The dense CSM was likely formed by the interaction of the slow but
dense red supergiant (RSG) wind with the fast, low-density, Wolf-Rayet
(W-R) or blue supergiant (BSG) wind starting some $10^3$ to $10^4$ yr
before the explosion.
Even when starting with spherically symmetric winds, factors such as
the magnetic field, turbulence within the bubble, presence of a binary
companion, and hydrodynamic, magneto-hydrodynamic and ionization front
instabilities \citep[see][and references therein]{Dwarkadas2022}, can
all lead to significant departures from sphericity. These can be
exacerbated by the presence of an asymmetric surrounding medium,
giving rise to a complex morphology.

\section{Total Flux Density Observations with MeerKAT, ATCA and ALMA}
\label{sobflux}

\subsection{MeerKAT Observations}
\label{sMeerKAT}

We observed SN~1996cr with MeerKAT\footnote{Operated by the South
African Radio Astronomy Observatory (SARAO).} on 2018 May 4 (proposal
code SCI-20180222-MB-01; we give the midpoint dates) and also used
archival MeerKAT observations from 2018 July 20 (code
SSV-20180428-FC-01), both
at a central frequency of 1.28 GHz with a bandwidth of 856 MHz split
into 4096 channels.  The data were reduced using a combination of the
OxKAT scripts \citep{Heywood2020_Oxkat} and manual reduction using the
Common Astronomy Software Applications package
\citep[\textsc{CASA};][]{CASA_2022}.
To determine the flux density of the SN on the scale
of Stevens-Reynolds 2016 \citep{Partridge+2016},
as well as to calibrate the instrumental bandpass, we observed the
sources PKS J1939$-$6342 and PKS J0408$-$6544 (QSO B0408$-$65) on 2004
May 4, and 3C~286 on 2018 July 20. For phase and amplitude
calibration, we used the sources PKS J1424$-$4913 and PKS
J1619$-$8418. % both obs.
The data were self-calibrated in phase, but not amplitude. 

The entire primary-beam area was imaged, and SN~1996cr and the
Circinus Active Galactic Nucleus (AGN), as well as unrelated
background sources were deconvolved using the CLEAN algorithm
(\textsc{CASA}:{\tt tclean}).  SN~1996cr is well separated from the
AGN of the Circinus Galaxy.  We estimated the flux densities by
fitting a model consisting of an unresolved source\footnote{An
unresolved source in the images is an elliptical Gaussian with the
dimensions and orientation fixed to those of the restoring beam.} and
a baseline- or zero-level to a small region of the image around
SN~1996cr, with the former being interpreted as the SN\@, and the
latter as the diffuse emission from the galaxy.  Although this
estimate of the diffuse emission is approximate, its value was smaller
than our stated uncertainties, so its presence does not significantly
impact our flux density determination for SN~1996cr. For the 2018 July
20 observation with the full MeerKAT, the resolution was
$\sim$5\arcsec, SN~1996cr was 99 mJy, the image background rms level
was 19~\muJb, and the fitted zero level was 2\% of the peak brightness
of SN~1996cr.  For the 2018 May 4 observation, for which only 16
dishes were used, the corresponding values were 8\arcsec, 107 mJy,
98~\muJb, and $\sim$5\% of the SN~1996cr peak.
As well as the statistical uncertainty, our flux density uncertainties
include, and are dominated by, a 10\% systematic uncertainty due to
the flux-density bootstrapping \citep[see, e.g.][]{Driessen+2022}.
All flux density determinations are listed in Table~\ref{tfluxd}.

\subsection{ATCA Observations}
\label{sATCA}

We observed SN~1996cr in 2020 with the Australia Telescope Compact Array
(ATCA; Project Code C3323)\@.
The observations were reduced using \textsc{CASA}.  We again used the
flux-density scale of Stevens-Reynolds 2016 \citep{Partridge+2016},
and set the scale using observations of PKS J1939$-$6342.
\citep{Partridge+2016}.  The data were self-calibrated in phase, but
not amplitude, and images deconvolved using the CLEAN algorithm
(\textsc{CASA}:{\tt tclean}). The FWHM resolution ranged from
2.8\arcsec\ at 2.1 GHz to 0.3\arcsec\ at 34 GHz.  The flux densities
were measured from the resulting CLEAN images in the same way as
described above for MeerKAT, and our uncertainties include an 5\%
uncertainty in the flux-density bootstrapping. The systematic
uncertainty is the dominant contribution to the uncertainty in the
flux density of SN~1996cr at all frequencies except 9 GHz.  The values
are listed in Table \ref{tfluxd}.

\subsection{ALMA Observations}
\label{sALMA}

We observed SN~1996cr also with the Atacama Large Millimeter Array
on several occasions in 2019, as part of program
\#2018.1.00007.S (PI: F. Bauer)
at 108.0, 223.0, 339.9, and 465.5 GHz (bands 3, 6, 7, and 8,
respectively) on the dates given in Table~\ref{tfluxd} below.  The
observations were taken with the Atacama Compact Array of ALMA (ACA),
and centred on the SN position, with a total bandwidth of 7.5 GHz in
each frequency. The data were reduced using \textsc{CASA}. The flux
density scale at ALMA is set by observations of one of a set of
unresolved calibrator sources (``grid'' sources), referred ultimately
to solar system objects, and is accurate to $\sim$10\%
\citep{Francis+2020}.
Spectral cubes were inspected by eye to flag and remove any channels
affected by strong line emission (e.g.\ CO) associated with the disk
of host galaxy.  Images were again made using the CLEAN algorithm
(\textsc{CASA}:{\tt tclean}), adopting Briggs weighting with {\tt
  robust = 0}.

At 108~GHz, the resolution of the ACA is 12\arcsec, and SN~1996cr and
the Circinus nucleus are $\sim$2 beamwidths apart, and have some
degree of overlapping emission.  There is also the possibility of some
diffuse emission from the galaxy at the location of SN~1996cr.  To
allow for these two factors, we determined the flux density of
SN~1996cr by fitting models to the region of the image near
SN~1996cr. We used three models: 1) an unresolved source for SN~1996cr
with a constant background level, 2) an unresolved source for
SN~1996cr, with a background level and slope, and 3) using a larger
fitting area, an unresolved source for SN~1996cr and an elliptical
Gaussian for the nucleus of the Circinus galaxy. We adopt the mean
over the three different models as the final flux density of
SN~1996cr, and the standard deviation over the three models as an
estimate of the uncertainty in separating SN~1996cr from the
background, to obtain a flux density of $4320 \pm 1120 \; \mu$Jy for
SN~1996cr.  (The galaxy contribution at SN~1996cr's location was
smaller than the uncertainty at $\sim 400$~\muJb).  The image rms was
300 \muJb, and we take 10\% for the flux calibration uncertainty to
arrive at a final value of $4320 \pm 1230 \; \mu$Jy.  This value is given
in Table~\ref{tfluxd}.

At 223 to 465 GHz (ALMA bands 6, 7, and 8), the extended thermal dust
emission associated with a nearby spiral arm dominates, and we could
only obtain upper limits on the flux density of SN~1996cr, which are
also listed in Table~\ref{tfluxd}.

\section{MeerKAT, ATCA and ALMA results: Total Flux Densities and SED}
\label{sresflux}

\begin{table}
\begin{minipage}[t]{0.48\textwidth}
\caption{Total Flux Density Measurements}
\label{tfluxd}
\begin{tabular}{r l@{}c@{}l c}
  \hline
  Freq. & Date  & Age\tablenotemark{a} & Telescope & Flux density\tablenotemark{b} \\
  ~(GHz) &   & ~(days) &  & ~~(mJy) \\
  \hline 
  %  the ATCA values are "Finals" from sn1996cr_2020ATCA.log
   1.28 & 2018 May \phn4  & 8275  & MeerKAT-16\tablenotemark{c} & $107 \pm 11$\\
  %  MeerKAT 107 [From Leon] +-  6.1 [MFB: add Leon's 3 + 10% in quad
   1.28 & 2018 Jul 20    & 8531  & MeerKAT  & $99 \pm 10$ \\  % using 10% bootstrapping
   2.10 & 2020 Jan 31  & 8911  & ATCA &     $60.8 \pm 3.5$ \\  % beam(geo_mean) 2.8"
   5.50 & 2020 Jan 31  & ~~~\" & ATCA &     $33.8 \pm 1.9$ \\  % beam(geo_mean) 1.4"
   9.00 & 2020 Jan 31  & ~~~\" & ATCA &     $25.1 \pm 5.5$ \\
  18.00 & 2020 Jan 31  & ~~~\" & ATCA &     $18.4 \pm 1.0$ \\
  34.00 & 2020 Jan 31  & ~~~\" & ATCA &     $12.0 \pm 0.7$ \\  % beam 0.36"; corrected uncert
  108.00 & 2019 Mar 10  & 8791  & ALMA & \phn\phn$4.32 \pm 1.23$ \\  % MJD 58759
  223.00 & 2019 May \phn5   & 8640  & ALMA &  $<3.5$\tablenotemark{d} \\ % MJD 58608
  339.90 & 2019 Apr 12 & 8617  & ALMA &  $<4.5$ \\ % MJD 58585
  465.50 & 2019 Aug \phn6 & 8733  & ALMA &  $<22$ \\ % MJD 58701
\hline
\end{tabular}
\\
\tablenotetext{a}{The age is the time since the explosion, calculated
  assuming an explosion date of 1995 Sep.\ 7, and for the midpoint of
  the observations, see text \S\ref{sintro}.}
\tablenotetext{b}{Our standard errors include the image background
  rms values and a 5\% flux-density calibration error, added in
  quadrature.}
\tablenotetext{c}{These observations were made using only 16 antennas,
  before all 64 dishes of MeerKAT became operational.}
\tablenotetext{d}{For the ALMA observations at $\nu>200$~GHz,
  SN~1996cr was no longer distinguishable from the galaxy emission in
  the image, and we give as an upper limit on its flux density the
  typical brightness per beam of the galaxy at the location of
  SN~1996cr.}
\end{minipage}
\end{table}

\begin{figure}
\centering 
\includegraphics[width=0.48\textwidth]{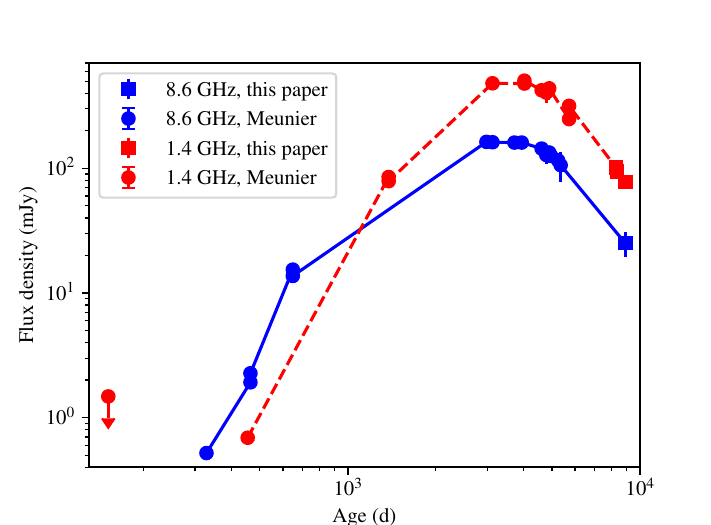}
%C  used sn1996cr_fluxplots v2.0; MKT data used 10% bootstrapping
\caption{The radio light curves of SN~1996cr at 8.6 and 1.4 GHz, with
  measurements from this paper and from \citet{Meunier+2013}.  The red
  points and dashed line show the 1.4 GHz light curve, while the blue
  points and solid line show the 8.6 GHz one. For the 1.4-GHz light
  curve we include values measured between 1.2 and 2.2~GHz, scaled to
  1.4 GHz assuming a spectral index of $-0.6$ (see \S~\ref{sresflux}),
  and for 8.4 GHz light curve we include values between 8 and 10 GHz,
  without any scaling.}
\label{flightcurve}
\end{figure}

We show the radio light curves of SN~1996cr at 8.6 and 1.4 GHz in
Figure~\ref{flightcurve}.  The flux density shows a steep decay since
the observations of \citet{Meunier+2013}, who had already suggested a
steepening of the decay near the end of their observations at $t \sim
6000$~d.  Between the last two measurements, the flux density decays
as $t^{-2.8 \pm 0.7}$ and $t^{-2.9 \pm 0.3}$ at 8.6 and 1.4 GHz,
respectively.

\begin{figure}
\centering 
\includegraphics[width=0.48\textwidth]{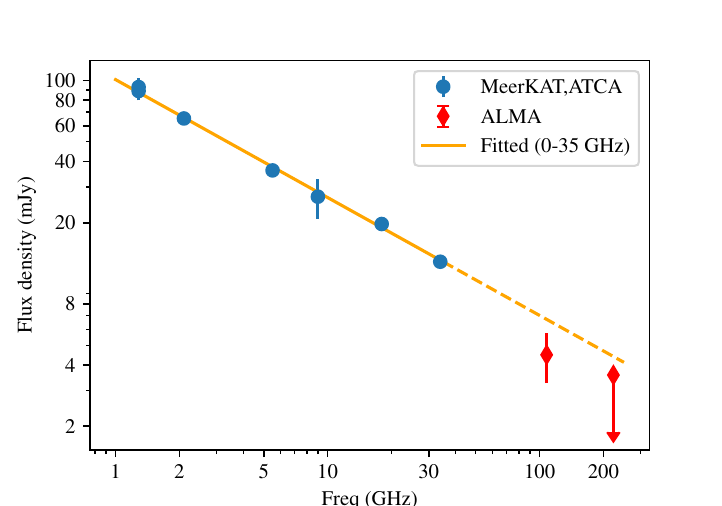}
%C  made with sn1996cr_fluxplots.py v2.1
\caption{The radio spectral energy distribution of SN~1996cr at $t =
  8700$~d (2019.5) from observations with MeerKAT, ATCA and ALMA\@.
  The orange line shows a power-law spectrum fitted by least squares
  for $\nu \leq 34$~GHz, which was $S(\nu) = 39.8 \; (\nu/{\rm 5 \,
    GHz})^{-0.580}$~mJy.  The flux density measurements have been
  scaled to a common time of $t = 8700$~d by assuming a decay rate of
  $S \propto t^{-2.85}$ as observed at 8.6 GHz, see
  \S~\ref{sresflux}.}
\label{fSED}
\end{figure}

We obtained new flux density measurements at frequencies between 1.28
and 465 GHz using MeerKAT, ATCA and ALMA (Table~\ref{tfluxd}) between
2018 and 2020 ($t = 8275$ and 8911~d).  We plot these flux density
measurements and a fitted power-law spectral energy distribution, SED,
in Figure~\ref{fSED}.
The SED shows no deviations from a power-law from 1.28~GHz up to the
highest ATCA measurement at 34~GHz, but the ALMA measurements at $\nu
> 100$~GHz show that the spectrum likely steepens somewhere between
34~GHz and 223 GHz.  We fit a power-law spectrum to $\nu \leq 34$~GHz
using weighted least squares.  Before fitting, we scaled all the
measurements to a common time of 8700~d by
assuming a decay $\propto t^{-2.85}$\@.  We obtain a fitted SED of
$S(\nu) = (39.8 \pm 1.0) (\nu/{\rm 5 \; GHz})^{-0.580 \pm 0.023}$ mJy,
with a $\chi^2 = 1.6$ for 5 degrees of freedom.

The spectral index between 34 and 108~GHz is $\alpha_{34\,\rm
  GHz}^{108\,\rm GHz} = -0.88\pm0.25$, and the probability that the
difference in $\alpha$ above and below 34~GHz, and thus the apparent
steepening above 35~GHz, are due to experimental error is
$\sim8$\%\footnote{This probability was derived numerically, the
distribution of $\alpha^{\rm 108 GHz}_{\rm 34 \, GHz}$ is somewhat
skew, with a tail to more negative values, so the probability of
getting a value $\geq -0.580$ is somewhat smaller than suggested by
Gaussian statistics.}.  Such a steepening might be expected to occur
because of synchrotron losses.  However, our results are not
conclusive on this point, and the degree of the steepening, as well as
the corner frequency, would have to be confirmed by better
high-frequency measurements.

\section{VLBI Observations}
\label{svlbiobs}

We observed SN~1996cr using the Australian Long Baseline Array, (LBA),
including the South African Hartebeesthoek antenna, at 2.3 GHz on 2020
Feb.\ 17 and at 4.8 GHz on 2020 March 3 (observing codes V253D and
V253E, respectively).  In both runs, the phased Australia Telescope
Compact Array (ATCA; $5 \times 22$~m diameter used for VLBI), Ceduna
(30~m), Hartebeesthoek (26~m), Hobart (26~m), Katherine (12~m), and
Parkes (64~m)
antennas took part, although no usable data were obtained from the
Katherine antenna.  At 2.3~GHz, the Tidbinbilla (DSS-36, 34~m),
Warkworth AU-Scope (12~m) and Yarragadee (12~m) antennas also
observed, although no usable data were obtained from either Yarragadee
or Warkworth.  At 4.8~GHz we also used the Warkworth 30~m and Mopra
(22~m) antennas.  In both cases, the observations were
phase-referenced to ICRF J142455.5$-$680758 (PKS J1424$-$6807), which
we will refer to as J1424$-$6807, and which is relatively unresolved
at both frequencies.

The calibration was carried out with NRAO's Astronomical Image
Processing System \citep[\textsc{AIPS};][]{Greisen2003}.  The initial
flux density calibration was done through measurements of the system
temperature at each telescope, and improved through self-calibration
of the data from the phase-reference source J1424$-$6807.

For the final image, we again used the CLEAN algorithm
(\textsc{AIPS}:{\tt imagr}), and we combined the 4.8 GHz and 2.3 GHz
data, so as to increase the \uv~coverage.  We scaled the nominal
visibility amplitudes of the 2.3-GHz data by $0.8\times$ so as to
match the visibility amplitude at the shortest baselines at both frequencies.  Since only
very partial $T_{\rm sys}$ measurements were available at 2.3 GHz, the
flux-density scale for the 2.3-GHz data is not well known.  If the
absolute amplitude calibration at 2.3 GHz were better, likely the
scaling between 2.3 and 4.8 GHz would be closer to the factor of 0.65
implied by the spectral index measured with ATCA and MeerKAT.

In addition to the 2020 VLBI observations described in this paper, we
had obtained earlier VLBI observations of SN~1996cr with the
Australian LBA, in 2007 at $t = 4307$~d and 22~GHz \citep[observing
  code VX013;][]{Bauer+2008}, and in 2013 at $t = 6553$ and 8.4~GHz
\citep[observing code V253C;][]{SNVLBI_Cagliari, SNVLBI_Crete}.  Both
these earlier LBA observations were phase-referenced to ICRF
J135546.6$-$632642 (PMN J1355$-$632), which is slightly closer on the
sky than J1424$-$6807\@.  ICRF J135546.6$-$632642 was, however
significantly resolved, and thus led to problems in the
phase-calibration and image reconstruction.
Because of these problems, and also because of the relatively 
poor \uv~coverage, there were significant ambiguities
in the image reconstruction, and we therefore do not use the images
from these epochs, but only use models fitted directly to the
visibility data.

Despite the difficulties in imagining, the 2013 image 
\citep{SNVLBI_Cagliari, SNVLBI_Crete} shows a
structure not dissimilar to the one from 2020 (see
Fig.~\ref{fvlbiimg}), which had much better \uv~coverage and was much
more reliably determined.

\section{VLBI Results: image}
\label{svlbiimg}

\begin{figure}
\centering 
%C SN96CR FCS 3.ICL001.7;
\includegraphics[width=0.48\textwidth]{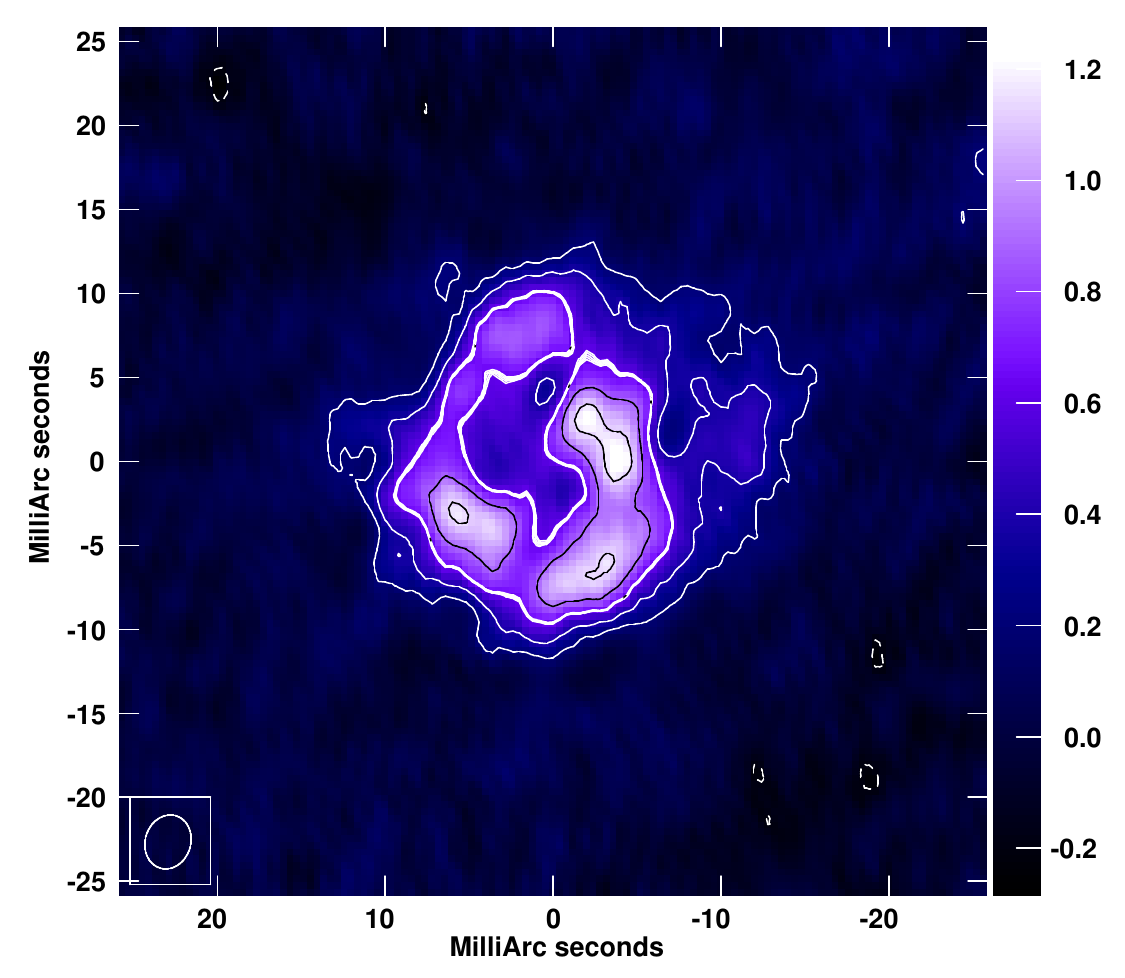}
\caption{The VLBI image of SN~1996cr at age $t \simeq 8959$~d, made
  with data from the Australian Long Baseline Array, combining the
  data at 5 GHz, observed 2020 March 28, and the amplitude-scaled data
  at 2.3 GHz, observed 2020 Feb.\ 17, (see text for details).  The
  restoring beam, indicated at lower left, was $3.54 \times 2.95$ mas
  at p.a.\ $-15$\arcdeg\ (FWHM).  With the combined data, we are
  sensitive to angular scales from 1~mas to 330 mas.  Both the
  colourscale, labelled in m\Jb, and the contours show brightness.
  The contours are drawn at $-16$, 16, 30, {\bf 50} (emphasized), 70
  and 90\% of the peak brightness, which was 1.26 m\Jb.  The rms
  background brightness was 72 \muJb.  North is up and east is to the
  left, and the origin is at the fitted position of the spherical
  shell model (see text, \S~\ref{ssize})}
\label{fvlbiimg}
\end{figure}

We show the VLBI image of SN~1996cr, made from the combined 2.3 and
4.8 GHz VLBI data taken in 2020, in Figure~\ref{fvlbiimg}.  The age of
SN~1996cr was 8929~d for the 2.3 GHz observations, and 8959~d for the
5 GHz observations taken around a month later.  Since the fractional
difference in age is small, and 5-GHz data dominate, we use 8959~d as
the representative age for the image. The combined 2.3 and 5 GHz
observations covered a range in \uv~spacing from 620 K$\lambda$ to 170
M$\lambda$, therefore we have some sensitivity to angular scales from
330~mas down to 1 mas.

\section{VLBI Results: size and expansion velocity}
\subsection{Size of the emission region}
\label{ssize}

Since SN~1996cr seems to have an approximately shell-like morphology,
we fit a geometrical model directly to the visibilities to determine a
precise value for the angular size.  The model we use is the same one
used for other SNe such as SN~1993J and SN~2014C
\citep[see][]{SN93J-3, SN2014C_VLBI}, and consists of the Fourier
transform of the projection of an optically thin shell of emission.
The model is characterized by the inner and outer angular radii of the
shell, $\thin, \, \thout$, and the total flux density.
\citet{SN93J_Manchester} showed that, in the case of SN~1993J, the
results obtained through \uv~plane model-fitting are superior to those
obtained in the image plane.  As in the imaging, we used the square
root of the data weights in the fitting, which makes the results more
robust at the expense of some statistical efficiency.

The outer angular radius of the model, \thout, is most closely
identified with the forward shock, and also most reliably
determined by the data.  Our resolution is not sufficient to reliably
determine both \thin\ and \thout, and we therefore first fix the ratio
$\roi = \thout/\thin$ at 1.25, as this value has been shown to be
appropriate in the case of SN~1993J \citep{SN93J-3, SN93J-4}. For the
case of a simple CSM structure and a non-magnetic shell, similar
values were also seen in numerical simulations \citep{JunN1996a},
although we note that the CSM structure in SN~1996cr is clearly not
simple.  Nonetheless, the fitted value of \thout\ is only weakly
dependent on the assumed value of \roi\ (with the fitted outer radius
varying on the order of $\sqrt{\roi}$).

Although our image (Fig.~\ref{fvlbiimg}) was made by combining the
scaled 2.3-GHz data with the 5-GHz data, it was not possible to
combine the two frequencies for model-fitting.  The 5-GHz data are much
more constraining, therefore we did the model-fitting only on the 5-GHz
VLBI data.

Our best-fit model had an outer angular radius of $\thout = 9.27$~mas,
with a statistical uncertainty of 0.04 mas, or $<0.5$\%.  However, as
we found for SN~2014C, the statistical uncertainty is dominated by the
systematic one.  We follow the same procedure we used in
\citet{SN2014C_VLBI} to estimate a systematic uncertainty, and again
include three contributions in our final standard error, added in
quadrature

The first contribution was estimated using jackknife re-sampling
\citep{McIntosh2016}.  Specifically, we dropped the data from each of
the antennas in the VLBI array in turn and calculated $N_{\rm antenna}
= 7$ new estimates (only 7 antennas contributed data to the final
results) of the fitted size, and the scatter over these 7 values
allows an estimate of the uncertainty of the original value that
included all antennas.  We obtained a jackknife uncertainty of
0.42~mas.  This is much larger than the purely statistical one because
the errors in the visibilities are dominated by residual calibration
errors, and perhaps departures of the true morphology from the assumed
model, both of which are strongly correlated from one visibility
measurement to the next, rather than purely random and uncorrelated
noise.

To determine the uncertainties due to the inexactly-determined antenna
gains, we performed a Monte-Carlo simulation by randomly adjusting the
antenna gains by an rms of 10\%. By subsequently refitting the size,
we obtained an rms scatter of the outer angular radius of 0.45 mas.
Adding these uncertainties in quadrature we obtain a final value for
the fitted outer angular radius of $9.27 \pm 0.60$~mas.  
This corresponds to a linear radius of $(5.1 \pm 0.3) \times
10^{17} \; (D / {\rm 3.7 \, Mpc})$~cm.

Although the optically-thin spherical shell model provides a
reasonable fit to the data, and a consistent way of evaluating the
expansion as a function of time, the morphology in the image in
Fig.~\ref{fvlbiimg} shows some deviations from that model.  As an
alternate, model-independent, means of estimating the mean radius, we
turn to \thfl\ \citep[introduced in][]{SN86J-2}, which is the mean
angular radius which encloses 90\% of the total flux density in the
image\footnote{In \citet{SN86J-2}, we also used the radius of the
lowest reliable contour in the image.  For SN~1996cr, the lowest
reliable contour was 16\%, and \thfl\ is virtually identical to the
angular radius of the 16\% contour in Fig.~\ref{fvlbiimg}, so we
consider only \thfl\ in what follows.}.  The \thfl\ radius is defined
as the square root of the area of the relevant contour in the image
divided by $\pi$.  Unlike the radius, \thout, of the model fitted
directly to the visibilities, \thfl\ is dependent to some degree on
the size of the convolving beam.  However, if the actual angular
radius is larger than the beam, the dependence is weak.

Measuring on the image of Fig.~\ref{fvlbiimg}, we find \thfl\ = 11.9
mas.  A test shows that on a spherical shell model, convolved with our
beam, \thfl\ is $1.02\times \thout$\@.  For SN~1996cr, we find, by
contrast, that \thfl\ is 28\% larger than \thout.  In other words,
SN~1996cr has faint radio emission that extends beyond what is
expected from a pure spherical-shell morphology.  This can be seen in
Fig.~\ref{fvlbiimg}, where, particularly to the west-northwest (WNW),
there is emission well beyond the bright ridge, which in a spherical
shell is approximately located near the \thin.

\begin{figure}
\centering 
\includegraphics[width=0.48\textwidth, trim=0.2in 0 0.2in 0, clip]{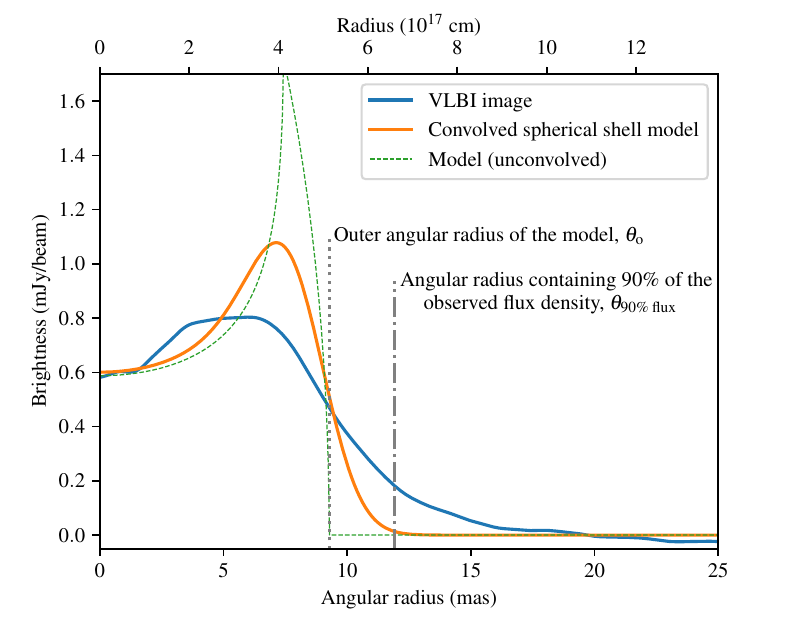}
%C made w/ radialprof.py v1.0. 
\caption{A radial profile of the brightness of SN~1996cr, derived from
  the image in Fig.~\protect\ref{fvlbiimg}, at age $t \simeq
  8959$~d\@.  The bottom horizontal axis shows angular radius, while the
  top shows linear radius (for $D = 3.7$~Mpc). The blue solid line shows
  the observed radial profile.  The orange solid line shows the
  profile of the fitted optically-thin spherical model, which has
  \thout = 9.27~mas and \roi = 1.25, convolved with our restoring beam
  ($3.54 \times 2.95$~mas at p.a.\ $-15\arcdeg$).  The dashed green
  line shows the profile of the model before convolution with the
  restoring beam. The grey dotted and dash-dotted vertical lines show
  the outer angular radius of the fitted model and the observed
  \thfl\ (the angular radius containing 90\% of the observed total
  flux density), respectively.}
\label{fradprof}
\end{figure}

We show a radial profile of brightness for SN~1996cr at $t = 8959$~d
in Figure~\ref{fradprof}, along with the corresponding profile
for the fitted model.  
SN~1996cr shows an excess of emission outside the fitted outer radius,
\thout, implying that a significant amount of radio emission occurs
outside the model's outer radius, thus at angular radii larger than \thout,
consequently also implying velocities higher than that derived from
\thout.  The maximum in the radial profile is flatter and more
extended than that in the model, suggesting that the radio emission is
more spread out radially than would be expected from our shell model,
which has \roi = 1.25.

The ridge line (or locus of brightest emission) in Fig.~\ref{fvlbiimg}
appears elliptical, with an axis ratio of $\sim$1.5:1.  If
interpreted as a ring or equatorial-belt structure, rather than a
complete spherical shell, this axis ratio suggests an inclination
angle with respect to our line of sight, $i = 48\arcdeg$.
This value of $i$ is close to that of 55\arcdeg\ estimated from the
X-ray line profiles by \citet{Quirola-Vasquez+2019}.  A CSM density
distribution that was high along the minor axis direction in the image
and low along the major axis one could produce such ellipticity from
an initially isotropic expansion.  Such a CSM density distribution is
unlikely to be due to gradients in the ISM density, but could easily
be created by an axisymmetric structure in either the ejecta or the
CSM\@.  This structure might be produced, for example, by a stellar wind with a
polar-angle dependent flow speed or mass-loss rate.  The VLBI image
further suggests a possible ``blow-out'' region to the WNW.

We find therefore that the VLBI image suggests somewhat anisotropic
expansion, which is also suggested by the X-ray line profiles
\citep{Quirola-Vasquez+2019} and the complex optical Oxygen-line
emission \citep{Bauer+2008}.  The furthest angular extent from the
fitted centre position in the VLBI image is $\sim$15~mas
(Figs.\ \ref{fvlbiimg}, \ref{fradprof}), corresponding to a linear
extent of $\sim 8\times 10^{17} \, (D / {\rm 3.7 \, Mpc})$ cm.

\subsection{Expansion Velocity}
\label{srexpansion}

\begin{figure}
\centering 
%C fexp.py  v2.3 (3.7 Mpc);
\includegraphics[width=0.48\textwidth, trim=0.1in 0.0 0 0.2in, clip]{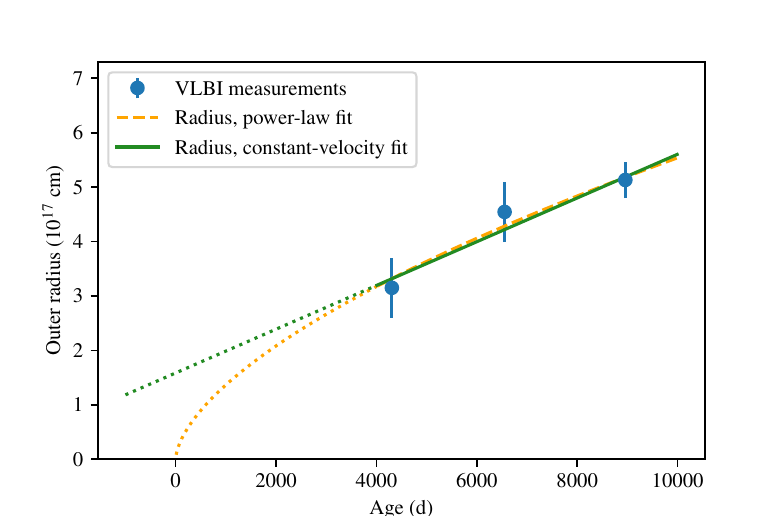}
\caption{The angular size of SN~1996cr.  We show the outer radii,
  \thout, from Table~\ref{tangs}, as determined by fitting a spherical
  shell model directly to the VLBI visibilities, along with their
  uncertainties, converted to a linear values for a distance of $D =
  3.7$~Mpc.  In addition, we show two fits to the outer radius values:
  first a power-law fit, of the form $r \propto t^m$, with $m = 0.608
  \pm 0.139$ (orange dashed line), and the second a constant velocity
  fit of the form $r = (t - t_0) \times\, v$, with $v = 4650 \pm
  1060$~\kms\ (green solid line).
  % [3.7 Mpc value for v];  see python chunk in this paper for soln
  %  Finally, with the scale on the right axis, we also show the velocity
  %corresponding to the power-law fit.
  Extrapolations beyond the range of VLBI measurements are given with
  respective dotted lines.}
\label{fexp}
\end{figure}

As mentioned in \S~\ref{svlbiobs}, we had also obtained earlier VLBI
observations of SN~1996cr at $t = 4307$~d (2007 June 24) and 22~GHz,
and again at $t = 6553$~d (2013 August 17) and 8.4~GHz
\citep[see][]{Bauer+2008, SNVLBI_Cagliari, SNVLBI_Crete}.  We fitted
the same spherical shell model to the visibilities of these earlier
observations to obtain angular size estimates also for these two
epochs.  The values we obtained were \thout = 5.69~mas at $t = 4307$~d
(2007) and 8.21~mas at $t = 6553$~d (2013).  Uncertainties are hard to
estimate since they will be dominated by the effects of the poor phase
calibration, but we estimate that a value of $\pm 1$~mas should
encompass a reasonable uncertainty range.  We tabulate the angular
sizes in Table~\ref{tangs} and plot the values in Figure~\ref{fexp}.

\begin{table}
\begin{minipage}[t]{0.48\textwidth}
\caption{Angular Size Estimates}
\label{tangs}
\begin{tabular}{l c l c}
  \hline
  Date  & Age\tablenotemark{a} & \thout\ (fitted shell model)\tablenotemark{b}
  & \thfl\tablenotemark{c} \\
        & (d)                  & \hspace{8mm}(mas) & (mas) \\
  \hline
  2020 Mar 28  & 8959  & \hspace{6mm}$9.27 \pm 0.60$  & 11.9 \\
  2013 Aug 17  & 6554  & \hspace{6mm}$8.21 \pm 1.00$\tablenotemark{d} \\
  2007 Jun 24  & 4307  & \hspace{6mm}$5.69 \pm 1.00$\tablenotemark{d} \\
  \hline
\end{tabular}
\tablenotetext{a}{The age is the time since the explosion, calculated
  to the midpoint of the observations and assuming an explosion date
  of 1995 Sep.\ 7, see text \S\ref{sintro}.}
\tablenotetext{b}{The outer angular radius of the fitted model of the projection
  of an optically thin, uniform spherical shell of emission.}
\tablenotetext{c}{The mean angular radius of the solid angle which
  encloses 90\% of the total flux density in the image.  Due to the
  unreliability of the images for 2001 and 2010 it was not possible
  to determine \thfl\ for those epochs.}
\tablenotetext{d}{Uncertainty is estimated.}
\end{minipage}
\end{table}

We fitted two expansion functions to our three values of the angular
radius (Table~\ref{tangs}) by least squares.
The first
function is a power-law, with
\begin{multline}
r(t) = (4.46 \pm 1.59) \times 10^{17} \times \\
  (\frac{t}{7000 \; {\rm d}})^{(0.608 +- 0.139)} \; (\frac{D}{\rm 3.7 \, Mpc})\; {\rm cm.}
\end{multline}
The second is a constant velocity fit, with
\begin{multline}
  r(t) = (2.9 \pm 1.3) \times 10^{17} + \\
  (4.0 \pm 0.9) \times 10^{13} 
  (\frac{t - 7000}{\rm d}) \; (\frac{D}{\rm 3.7 \, Mpc}) \; \rm{cm,}
\end{multline}  
corresponding to a velocity of $4650 \pm 1060 \; (D / {\rm 3.7 \,
  Mpc})$~\kms.
The two fitted functions are also plotted in Fig.~\ref{fexp}.  The
measurements do not distinguish between these two alternate functions.

At $t = 8959$~d, the fitted outer angular radius corresponds to $(5.1
\pm 0.3) \times 10^{17} \times (D / {\rm 3.7 \, Mpc})$~cm.  This
radius corresponds in turn to an average expansion since the
explosion, at $t_0$, of $(6630 \pm 430) \; (D / {\rm 3.7 \, Mpc}$
\kms\ (where the uncertainty includes the contribution of the
$\sim$200~d uncertainty in $t_0$).
However, the radio emission extends beyond this average outer radius,
particularly to the WNW, so in that direction, the mean expansion
velocity since $t_0$ is $\sim 10^4$~\kms,
50\% larger than the value corresponding to the outer radius of the
fitted shell model.

The current velocity, estimated from our constant velocity fit between
4307 and 8959~d, is ($4650 \pm 1060) \; (D / 3.7 \, \rm Mpc)$~\kms,
and is clearly lower than the average velocity since $t_0$.
Therefore, the expansion early on in SN~1996cr's history must have
been more rapid than it was since 4307~d.

We estimate the current velocities corresponding to \thfl\ and to the
maximum extent of the radio emission to the WNW by scaling the values
obtained from \thout\ and the fitted shell model, because we cannot
reliably estimate those values directly from the 2007 and 2013 images.
We obtain current velocities of $\sim$5970~\kms\ for \thfl\ and
7000~\kms\ to the WNW, both scaling as $D/{\rm 3.7\,Mpc}$.

In summary, a representative value for the current velocity of the
forward shock is $(4650 \pm 1060) \; (D / {\rm 3.7 \, Mpc})$~\kms, but,
judging from the extended emission to the WNW, a range of current
velocities extending up to $\sim 7000 \; (D / {\rm
  3.7\,Mpc})$~\kms\ seems to be present.

\section{Discussion}
\label{sdiscuss}

SN~1996cr, in the Circinus Galaxy, had the highest radio brightness of
any spectroscopically-identified SN\@.  We obtained new radio
observations of SN~1996cr.  Specifically, we obtained VLBI
observations, allowing us to resolve the radio emission spatially at
three epochs (4300~d to 9000~d), allowing us to infer the expansion
velocity.  In addition, we obtained broadband total flux density
measurements at $\sim$8600~d using MeerKAT, ATCA and ALMA, allowing us
to monitor the evolution of the radio SED.

\subsection{Spectral Index}
\label{sdiscussalpha}

At age $t = 8700$~d (2019.5), SN~1996cr's SED shows an unbroken
power-law down to the values measured by MeerKAT at 1.284 GHz.
The synchrotron self-absorption frequency must therefore be
$<1.28$~GHz. Taking the average radius of $5.1\times10^{17}$~cm
(\S~\ref{ssize}), and assuming the power-law spectrum with $S \propto
\nu^{-0.59}$ extends down to the synchrotron self-absorption
frequency, $\nu_{\rm SSA}$, we can calculate \citep{Chevalier1998}
that $\nu_{\rm SSA} = 176$ MHz, and thus well below our observations.
Could we observe the SN at this frequency? The best telescope at this
frequency would be the Murchison Widefield Array (MWA), but it would
not be possible to separate SN~1996cr from the Circinus Galaxy
nucleus, only $\sim$23\arcsec\ away or from the diffuse emission from
the galaxy.

The SED was relatively steep at $t = 397$~d but subsequently flattened
with time, with $\alpha$ increasing at $+0.017 \pm 0.007$~year$^{-1}$
to $\alpha = -0.76$ by $t=5370$~d \citep{Meunier+2013}.  Our value of
$\alpha = -0.580 \pm 0.023$ at $t = 8700$~d is entirely consistent
with that increase in time, despite the increase in the rate of flux
density decay with time since $t = 5307$.

\subsection{Radio Spectral Luminosity and its Decay}
\label{ssdecay}

We measured a total 9-GHz flux density, $S_{\rm 9\,GHz}$, at $t =8911$~d
(2020), of $25.1 \pm 5.5$~mJy, corresponding to a spectral luminosity,
$L_{\rm 9 \, GHz}$ of $(4.1 \pm 0.9) \times 10^{26} \; (D / {\rm 3.7 \,
  Mpc})^2$ erg~s$^{-1}$ Hz$^{-1}$.
Few SNe have been observed at such large $t$, but that value of
$L_{\rm 9 \, GHz}$ falls within the relatively large range of values one
might extrapolate for other Type IIn SNe at $t = 8911$~d
\citep{RadioLumFn}. In particular, SN~1996cr's $L_{\rm 9 \, GHz}$ at $t
\sim 9000$~d is about twice as high as that of SN~1986J, and we note
that the latter also shows a steep dropoff of radio luminosity at late
times \citep{SN86J-4}.

Over the last two years, SN~1996cr's flux density has decayed rapidly,
with $S_{\rm 9 \, GHz} \propto t^{-2.8 \pm 0.7}$. The two MeerKAT
measurements in 2018 at 1.284 GHz give a consistent decay rate, albeit
with large uncertainty, of $S_{\rm 1.3 GHz} \propto t^{-2.6 \pm
  2.3}$\@.
In the standard, self-similar model of SN evolution, with both ejecta
and CSM density profiles being power-laws in radius, there is
no solution with the flux density decay as steep as this,
however, steep flux density decay generally corresponds to
a steep CSM density profile
\citep[see, e.g.][]{FranssonLC1996}.  While SN~1996cr's evolution is
clearly not self-similar, the steep observed decline in the flux
density may nonetheless suggest, at the present radius of $\sim
5\times10^{17}$~cm, a relatively steep CSM density profile, with $\rho
\propto r^{-s}$ with $s>2$, which is steeper than a wind with constant
mass-loss rate and velocity, which would have $s=2$\@.  This picture
is the opposite of what we found for SN~2014C, where the SN shock also
interacted with a region of dense CSM, but we found that the CSM
density outside of the dense region likely had a flat profile with $s
< 2$ \citep{SN2014C_VLBI2}.

\subsection{Morphology}
\label{sdiscusmorph}

Our VLBI image of SN~1996cr at $t \sim 8959$~d (Fig.~\ref{fvlbiimg})
shows a morphology resembling that of an optically-thin spherical
shell seen in projection. Such a morphology is similar to that of the
other SNe that have so far been reliably resolved
(\citealt{SNVLBI_Cagliari,BartelKB2017}; for example, see
e.g.\ \citealt{SN93J-3, Marcaide+1995b} on SN~1993J).  A shell-like
morphology is also expected on theoretical grounds
\citep[e.g.][]{ChevalierF2017}.

However, SN~1996cr shows notable departures from a circularly
symmetric morphology on the sky.  The ridge line, or locus
of brightest emission, appears to be somewhat elliptical, and emission
of lower surface brightness extends well beyond the ridge-line,
especially to the WNW\@.  The elliptical ridge-line suggests a
circular structure tilted to the plane of the sky by $\sim$48\arcdeg.

\citet{Quirola-Vasquez+2019} found that the X-ray line profiles also
suggested complex morphology for SN~1996cr, with neither a spherical
shell nor a simple ring-like geometry being compatible with the
observed profiles.
In particular, \citet{Quirola-Vasquez+2019} suggested that there are
two X-ray emitting components: a lower-temperature one with $T \simeq
2$~keV with a wide angle of outflow, and one with a higher $T \simeq
20$~keV and a narrower angle of outflow.  They associate the hotter,
narrow-angle outflow with a reflected shock, formed when the ejecta
impacted on the dense CSM region, currently propagating inward (in
a co-moving frame) into the already shocked CSM interior to the
forward shock, further heating this material and thus producing the
high temperatures.

\subsection{Expansion Velocity}
\label{sdiscussv}

We determined the expansion velocity of SN~1996cr by tracking the
evolution of the radius between our three VLBI epochs.  We obtained
mean radii of SN~1996cr by fitting an optically-thin spherical shell
model, which can account for the bulk of the emission, to our VLBI
data, and we took the outer radius of this fitted model to be a
representative mean value for the radio emission region and location
of the forward shock.  At $t = 8859$~d, this value was $(5.1 \pm 0.4)
\times 10^{17} \; (D / {\rm 3.7 \, Mpc})$~cm, corresponding to a mean
velocity since the explosion of $(6700 \pm 430) \; (D / {\rm 3.7 \,
  Mpc})$ \kms.  This velocity is comparable to, albeit slightly lower
than that seen for other Type II SNe at similar ages: for example,
SN~1993J had a mean velocity of 11,300~\kms\ at $t=3164$~d
\citep{SN93J-2}, which, extrapolating the deceleration, would equal
9700~\kms\ at $t = 8959$~d;
while SN~1986J had one of 8790~\kms\ at $t=8333$~d \citep{SN86J-2}.

By using also the radii determined from our VLBI measurements from
2007 and 2013, we found that the present expansion velocity of
SN~1996cr, again as determined from the fitted spherical shell model,
was roughly constant at $(4650 \pm1060) \; (D / {\rm 3.7 \, Mpc})$~\kms\ over
the period $4307 \leq t \leq 8859$~d.

Our measurements clearly show that the forward shock has been
decelerated since the explosion: the explosion date around 1995
Sep.\ \citep[see, e.g.][]{Bauer+2008} requires that the velocity
before our first VLBI measurement at $t < 4307$~d must have been
higher at $\gtrsim$ 11,000~\kms\ (Fig.~\ref{fexp}).

\citet{Quirola-Vasquez+2019} determined velocities from the X-ray line
profiles, and at $t \simeq 4900$~d, found maximum velocities of
$\sim$5000~\kms\ for elements such as Si, but only
$\sim$3000~\kms\ for Fe at $t \simeq 4900$~d.  The velocities seen in
the X-ray, being lower than those seen in the radio, are consistent
with the expectation that the X-ray emission arises near the reverse
shock, or near a reflected shock, formed when the forward shock first
encountered the dense CSM at $t \sim 500$~d, since the reverse and
reflected shocks are generally expected to have a lower velocity than
the forward one.

There is a wealth of evidence that the expanding shock in SN~1996cr
first moved rapidly through a relatively low density medium
immediately surrounding the progenitor, before impacting on dense CSM
at radius $\sim \! 10^{17}$~cm at around $t \sim 500$~d after the
explosion \citep{Bauer+2008, DwarkadasDB2010, Meunier+2013,
  Quirola-Vasquez+2019}.
%CSM shell.
Our earliest VLBI measurement at $t = 4307$~d, however, was
considerably later than this impact. Indeed, in the simulations of
\citet{DwarkadasDB2010}, the forward shock impacts on a shell of dense
CSM shell, then passes through and exits the shell again at $t \sim
2500$~d, all before the first VLBI measurement.  In those simulations
the shock velocity remains approximately constant at $v \sim
4650$~\kms\ after $t \sim 3000$~d.
This behaviour corresponds well to the VLBI measurements of the
expansion.

The shock is thought to have impacted on the dense CSM at around $t
\sim 500$~d. Extrapolating our constant-velocity fits to the expansion
curve back towards that time, we obtain a radius of $(1.8 \pm 0.6)
\times 10^{17} \; (D / {\rm 3.7\,Mpc})$~cm.
The VLBI measurements therefore suggest an outer radius of $\sim 1.8
\times 10^{17}$~cm at the time of the impact of the shock on the dense
CSM\@.  This value is somewhat larger than has been used in the
modelling, e.g.\ $10^{17}$~cm in \citet{DwarkadasDB2010} and
\citet{DeweyBD2011}, suggesting a more rapid expansion before the
impact on the dense CSM.

Given that the VLBI image (Fig.~\ref{fvlbiimg}) shows some significant
departures from a spherical-shell morphology, a range of expansion
velocities for the forward shock will be present at any one time.  In
particular the radio emission beyond the boundary of the fitted
spherical shell suggests expansion velocities, for $4307 \leq t \leq
8859$~d, of up to $\sim$7000~\kms\ to the WNW (see
\S~\ref{srexpansion}).

Although the limited dynamic range and resolution of the VLBI image do
not allow a conclusive determination of the three-dimensional source
structure, the morphology could perhaps be interpreted in terms of a
ring or equatorial-band-like radio-bright region, tilted to the line
of sight.  This radio-bright region is likely produced by a
similarly-shaped region of higher density in the CSM\@.

There appears to be some similarity between the radio morphology of
SN~1987A and SN~1996cr. We may therefore go a step further and assume
that there is a similarity in their formation mechanisms.  In
SN~1987A, the brighter radio emission is caused by the interaction of
the shock with the narrow waist of an hourglass or peanut-shaped
cavity in an \HII\ region into which the SN expands, which results in
a equatorial-belt like region of denser CSM and consequently brighter
radio emission \citep{Orlando+2019}.  Note that in SN~1987A's case
there is also a set of three rings of even higher density, however,
these are not thought to affect the radio brightness greatly.  The
progenitor of SN~1987A is presumed to have had a binary companion, and
it is thought that binary interaction shaped the cavity in the
\HII\ region \citep[and also the dense rings; see e.g.][]{Potter+2014,
  Orlando+2015,Orlando+2019}.

We may then postulate, on the basis of the morphological similarity,
that SN~1996cr's progenitor star also had a companion, and that binary
interaction was also the cause of the global asymmetry.  SN~1996cr's
progenitor was constrained to be a W-R or BSG star by
\citet{DwarkadasDB2010}.
Similar to the scenario in SN~1987A, it is the interaction of the wind
from the progenitor immediately prior to the explosion, that is when
the progenitor was in the W-R or BSG epoch if its evolution, with that
from a previous, likely RSG epoch, that leads to the
formation of a wind bubble around the star. Asymmetries in the RSG
wind would then lead to a global asymmetry in CSM structure, which
would be reflected in the radio and X-ray profiles of the SN\@.  Note
that simulations have shown that there are considerable instabilities
in the wind-blown bubbles around massive stars, which lead strong
departures from spherical symmetry for the CSM \citep[see,
  e.g.][]{Dwarkadas2022}, although such instabilities are more likely to
produce a clumpy structure than the global asymmetries seen in the
radio emission.  Such instabilities may, however, enhance any global
asymmetries present in the wind.

The forward shock is slowed significantly where it impacts on the
denser CSM, but continues to expand more rapidly along the directions
where the CSM is less dense.  In SN~1996cr's case, the forward shock
encountered the dense CSM first at $t \sim 500$~d at a radius of $\sim
\! 1.8\times 10^{17}$~cm (as opposed to $\sim \! 3 \times 10^{17}$~cm for
SN~1987A).
In this picture, that denser CSM would be a ring or equatorial belt,
perhaps the narrow waist of an hourglass- or peanut-shaped cavity (as
in SN~1987A).  The interaction would then produce the tilted-ring
morphology of the ridge-line in the VLBI image.  Other portions of the
forward shock, however, moving along the axial direction, would
continue moving more rapidly.  It seems also that some departures from
cylindrical symmetry are required to explain the asymmetry in the VLBI
image, which has more extended emission to the WNW than to the ESE\@.
Such an asymmetrical ``blowout'' is seen in the supernova remnant SNR
G0.9+0.1 where the outflow is likely due to the pulsar embedded in the
nebula, but is considerably more pronounced on one side than the other
\citep{Heywood+2022}.  Such an asymmetry could perhaps also explain
the asymmetries seen in the X-ray line profiles.

\section{Conclusions and Summary}
\label{sconc}

We obtained VLBI observations of SN~1996cr, and for the first
time reliably imaged the radio emission.  We have also obtained
total flux density observations allowing us to monitor the evolution
of the radio spectral energy distribution.  SN~1996cr is thought to
have exploded in a low-density region, but then the shock interacted
with a massive shell in the CSM after $t = 1\sim2$~yr.  We found the
following:

\begin{trivlist}
  
\item{1.} Our new VLBI observations show a structure which is
  relatively circular in outline.  Within this outline, the brightest
  emission is somewhat elliptical, but generally near the outside edge
  (edge-brightened), except on WNW, where fainter emission extends
  well beyond the bright ridge-line.  The morphology in the VLBI image
  is approximately consistent with the projection of a spherical
  shell, but the slight ellipticity of the ridge-line may suggest
  instead a circular structure, tilted to the plane of the sky.

\item{2.} At $t = 8859$~d, the average outer radius of the radio
  emission, likely close to that of the forward shock, is $(5.1 \pm
  0.3) \times 10^{17} \; (D / {\rm 3.7\,Mpc})$~cm, as determined by
  fitting a spherical shell model to the radio emission.
  
\item{3.} The average expansion velocity since the explosion, again
  determined from the fitted spherical shell, is $6630 \pm 430 \;
  (D/{\rm 3.7 \, Mpc})$ \kms.  The average expansion velocity over the
  time of the three VLBI measurements ($4307 \leq t \leq 8959$~d, or
  2007 to 2020) was $(4650 \pm 1060) \; (D / {\rm 3.7 \, Mpc})$~\kms.
  This difference implies that the expansion had been significantly decelerated
  already before our first VLBI measurement at $t = 4307$~d\@.
  Although our data do not allow us to say whether this deceleration
  is ongoing, or whether it occurred mostly before $t = 4307$~d, the
  physical picture of the shock impacting on a dense CSM shell at $t
  \sim 500$~d suggests the latter.

\item{4.} The deviations of the observed radio morphology from that of
  a projected spherical shell suggest the presence of a range of
  expansion velocities for the forward shock, in particular, with some
  segments expanding at up to $\sim$7000~\kms\ during $4307 \leq t
  \leq 8959$~d.  The range of velocities seen in X-ray line profiles
  is a bit lower, consistent with the expectation that the X-ray
  emission comes from the reverse or reflected shocks interior to the
  forward one.

\item{5.} If we assume an approximately constant velocity expansion
  since $t \sim 500$~d, then the VLBI measurements suggest a radius at
  that time of $(1.8 \pm 0.6) \times 10^{17} \; (D / {\rm 3.7 \,
    Mpc})$~cm.  This radius therefore corresponds to the inner radius
  of the dense CSM region, and is somewhat larger than the $\sim
  1\times10^{17}$~cm that has been assumed in the hydrodynamic models
  so far, suggesting higher initial expansion velocities and a
  stronger deceleration upon interacting with the dense CSM shell, or
  an asymmetry in the ejecta/shell, or perhaps an incomplete shell.

\item{6.} SN~1996cr's radio flux density has decreased rapidly since
  2006, proportional to $t^{-2.9}$.  This could be due to a steep CSM
  density profile, with $s > 2$ where $\rho \propto r^{-s}$.

\item{7.} The spectral index of the radio emission between 1 and 34
  GHz at $t = 8700$~d (2019.5), was $\alpha = -0.580\pm 0.023$ with
  the spectrum over that range well described by a power-law.  The
  spectrum likely steepens above 35~GHz.

\item{8.} The spectrum below 35~GHz, which we found had $\alpha =
  -0.580 \pm 0.023$ at $t = 8700$~d, has continued to flatten with
  time since the value of $\alpha = -0.76$ reported at $t = 5370$~d by
  \citet{Meunier+2013}.

\end{trivlist}

\section*{Acknowledgements }

We thank Tony Foley (1957 to 2021) for carrying out the observations
of SN~1996cr with MeerKAT.  We thank the teams of the LBA,
MeerKAT and ATCA for their work to make the observations possible.
The Australia Telescope Compact Array is part of the Australia
Telescope National Facility (\url{https://ror.org/05qajvd42}) which is
funded by the Australian Government for operation as a National
Facility managed by CSIRO\@.  We acknowledge the Gomeroi people as the
traditional owners of the ATCA Observatory site.
the South African Radio Astronomy Observatory, a facility of
the National Research Foundation, an agency of the Department of
Science and Innovation.
This paper makes use of the following ALMA data:
ADS/JAO.ALMA\#2018.1.00007.S\@.  ALMA is a partnership of ESO
(representing its member states), NSF (USA) and NINS (Japan), together
with NRC (Canada), MOST and ASIAA (Taiwan), and KASI (Republic of
Korea), in cooperation with the Republic of Chile. The Joint ALMA
Observatory is operated by ESO, AUI/NRAO and NAOJ.
This research has made use of the NASA/IPAC Extragalactic
Database (NED), which is funded by the National Aeronautics
and Space Administration and operated by the California Institute
of Technology.
We have made use of NASA's
Astrophysics Data System Abstract Service. This research was supported
by both the National Sciences and Engineering Research Council of
Canada and the National Research Foundation of South
Africa. Additional funding for FEB was provided by ANID - Millennium
Science Initiative Program - ICN12\_009, CATA-Basal - FB210003 (FEB),
and FONDECYT Regular - 1190818 and 1200495.  VVD is supported by NSF
grant AST-1911061.

\section*{Data Availability Statement}

The raw data underlying this paper are available in the Australian
Telescope National Facility archive, \url{http://atoa.atnf.csiro.au},
the South African Radio Astronomy Observatory archive
\url{https://archive.sarao.ac.za}, and the ALMA archive
\url{https://almascience.nrao.edu/aq}.  Other data will be shared on
reasonable request.

\bibliographystyle{mnras}
\bibliography{mybib1,sn1996cr,sn2014c_temp2}

\label{lastpage}

\clearpage

\end{document}